# Is BERTopic Better than PLSA for Extracting Key Topics in Aviation Safety Reports?

Aziida Nanyonga
School of Engineering and Technology
University of New South Wales
Canberra, Australia
a.nanyonga@adfa.edu.au

Keith Joiner
School of Engineering and Technology,
University of New South Wales,
Canberra, Australia
k.joiner@unsw.edu.au

Ugur Turhan
School of Science
University of New South Wales
Canberra, Australia
u.turhan@unsw.edu.au

Graham Wild
School of Science
University of New South Wales
Canberra, Australia
g.wild@adfa.edu.au

***Abstract*—This study compares the effectiveness of BERTopic and Probabilistic Latent Semantic Analysis (PLSA) in extracting meaningful topics from aviation safety reports aiming to enhance the understanding of patterns in aviation incident data. Using a dataset of over 36,000 National Transportation Safety Board (NTSB) reports from 2000–2020, BERTopic employed transformer-based embeddings and hierarchical clustering, while PLSA utilized probabilistic modelling through the Expectation-Maximization (EM) algorithm. Results showed that BERTopic outperformed PLSA in topic coherence, achieving a C_v score of 0.41 compared to PLSA's 0.37, while also demonstrating superior interpretability as validated by aviation safety experts. These findings underscore the advantages of modern transformer-based approaches in analyzing complex aviation datasets, paving the way for enhanced insights and informed decision-making in aviation safety. Future work will explore hybrid models, multilingual datasets, and advanced clustering techniques to further improve topic modelling in this domain.**

## I. Introduction

The analysis of aviation safety reports is critical for identifying recurring issues and implementing measures to improve flight safety [1]. These reports, which often contain detailed narratives of incidents and accidents, provide a rich source of information for identifying trends, understanding underlying factors, and informing policy decisions. As the volume of aviation safety data continues to grow, advanced computational tools are required to efficiently extract meaningful insights from this textual data [2].

Topic modelling has emerged as a powerful technique for uncovering latent themes within large corpora of text. It allows researchers and practitioners to group related terms into coherent topics, facilitating a better understanding of the data. Over the years, several topic modelling techniques have been developed, ranging from traditional probabilistic approaches like Latent Dirichlet Allocation (LDA) and Probabilistic Latent Semantic Analysis (PLSA) to modern transformer-based methods like BERTopic [3, 4]. Each technique has its strengths and limitations, and its effectiveness often depends on the specific domain and dataset being analyzed.

PLSA, introduced by Hofmann (1999), was one of the first probabilistic approaches to topic modelling [5]. It assumes that each document is a mixture of topics, and each topic is a distribution over words. Despite its simplicity and historical significance, PLSA has several limitations, including scalability issues and challenges in interpreting its results when applied to large, complex datasets. Its reliance on a bag-of-words representation also limits its ability to capture contextual relationships between words, which are often crucial for understanding nuanced domains like aviation safety.

BERTopic, on the other hand, represents a significant advancement in topic modelling by leveraging transformer-based embeddings and clustering algorithms. This method integrates the contextual understanding of transformers with dynamic topic representation, enabling it to capture more meaningful and coherent topics. Unlike traditional methods, BERTopic can handle large datasets with high-dimensional textual data, making it particularly well-suited for domains with evolving language and diverse reporting styles, such as aviation safety [6].

In the context of aviation safety, where accurate identification of themes such as "engine failures," "runway incursions," or "pilot error" can have life-saving implications, selecting the right topic modelling approach is crucial. This study addresses the question: "Is BERTopic better than PLSA for extracting key topics in aviation safety reports?" To answer this, we compare these two techniques using reports sourced from the National Transportation Safety Board (NTSB). These reports are categorized under "Flight Operation Type" and provide a comprehensive overview of incidents and accidents within the aviation industry.

The contributions of this study are threefold: we provide a detailed comparison of BERTopic and PLSA, focusing on their ability to extract coherent and interpretable topics from aviation safety reports. We evaluate their performance using metrics such as topic coherence scores, scalability, and manual inspection by domain experts. Finally, we discuss the implications of our findings for the aviation industry and highlight potential areas for future research.

By exploring the strengths and limitations of BERTopic and PLSA, this paper aims to inform the choice of topic modelling techniques for analyzing aviation safety reports and other similar datasets. The findings are expected to contribute to the growing body of literature on topic modelling and its applications in critical domains.

The remainder of this paper is organized as follows: Section 2 reviews related work in topic modeling applications,

particularly in aviation safety. Section 3 outlines the methodology, including dataset details and evaluation metrics. Section 4 presents the results and discussion. Finally, Section 5 concludes with insights and future directions for research in this domain.

## II. Related Work

Topic modelling has become an essential technique in NLP for uncovering latent semantic structures in textual data. Numerous methodologies have been developed over the years, each with distinct underlying mechanisms and applications. This section provides an overview of relevant work on topic modelling, focusing on PLSA and BERTopic, and their applications in aviation safety and other domains.

PLSA, [5], is a foundational probabilistic approach to topic modelling. It models the co-occurrence of terms and documents using a latent variable representing topics. PLSA assumes that each document is a probabilistic mixture of topics, and a distribution over words characterizes each topic. Despite its simplicity and interpretability, PLSA is not without limitations.

Studies have highlighted its tendency to overfit, particularly on smaller datasets, as PLSA lacks a well-defined generative model for new documents [7, 8]. Furthermore, its reliance on the bag-of-words representation restricts its ability to capture semantic context, a critical limitation for analyzing complex domains such as aviation safety. However, PLSA has found application in various fields, including bioinformatics and information retrieval [9] showcasing its foundational importance.

BERTopic, a more recent advancement in topic modelling, integrates transformer-based embeddings with clustering techniques to generate dynamic and contextual topics [10]. By leveraging sentence embeddings from models such as BERT [11], BERTopic captures semantic relationships between words and phrases, offering improved coherence and interpretability compared to traditional approaches.

Research demonstrates BERTopic's superiority in domains requiring nuanced understanding, including healthcare [12], finance [13], and social media analysis [14]. Unlike PLSA, BERTopic adapts to evolving language patterns and effectively handles high-dimensional data. For instance, its application in analyzing incident reports from the aviation domain has shown promising results in identifying trends and recurring safety concerns [15].

The application of topic modelling to aviation safety is relatively nascent but growing. Aviation safety reports, such as those provided by the NTSB and Aviation Safety Network (ASN), contain unstructured narratives detailing incidents and accidents. These reports are invaluable for identifying systemic issues, such as mechanical failures or human errors, and for developing preventive strategies [16].

Previous studies have employed LDA and its variants to extract themes from aviation. However, LDA's reliance on simplistic text representations often results in less coherent topics. PLSA, similarly, has been used in the early analysis of safety reports but struggled with scalability and context-specific nuances [17].

Recent advancements, such as BERTopic, have demonstrated significant potential in this domain. Grootendorst's work [10] showed that BERTopic could extract actionable topics from complex datasets, including safety and risk analysis reports. Compared to PLSA, BERTopic offers better scalability and topic coherence, which are critical for analyzing aviation safety narratives that evolve.

Few studies have directly compared traditional probabilistic methods like PLSA with modern transformer-based techniques such as BERTopic. A study [18] conducted a comparative study on stock overflow datasets and found BERTopic to outperform PLSA in terms of topic coherence and interpretability. Similarly, another study observed that transformer-based models provided more relevant insights in customer call document analysis [3]. However, these comparisons are yet to be explored extensively within the aviation safety domain, leaving a gap that this study aims to address.

By building on these previous works, this study investigates the efficacy of BERTopic and PLSA in extracting meaningful topics from aviation safety reports, with a focus on coherence, relevance, and domain-specific applicability.

## III. Methodology

This section outlines the methodology used to evaluate the performance of BERTopic and PLSA in extracting key topics from aviation safety reports. The approach involves data collection and preprocessing, and implementation of the topic modelling techniques as shown in Fig. 1.

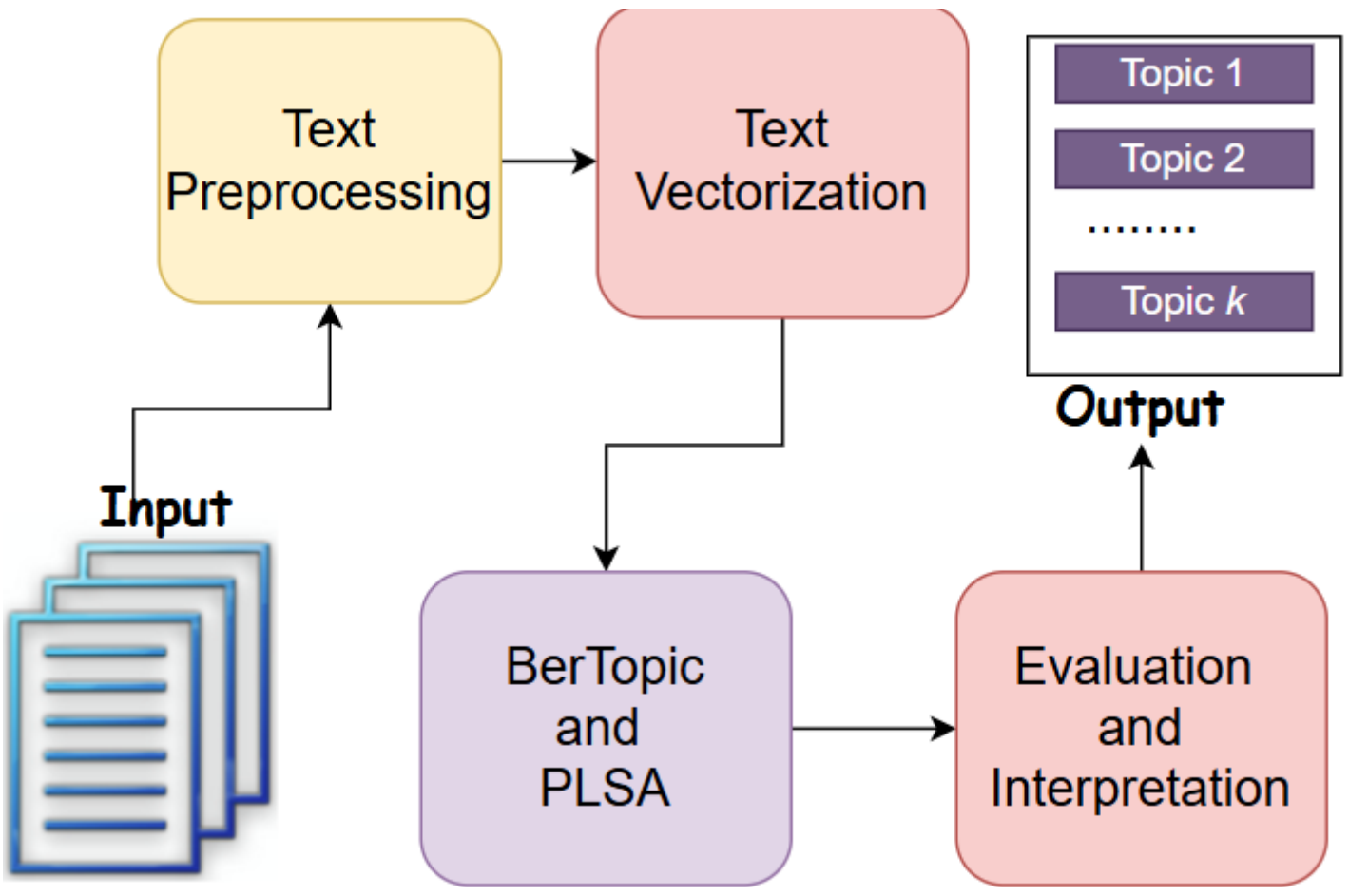

Fig. 1. Methodological framework

### A. Data Collection

Aviation incident and accident investigation reports used in this study were exclusively sourced from the NTSB spanning the years from 2000 to 2020. The dataset, comprising a collection of more than 36,000 records in JSON format, was obtained from https://www.ntsb.gov/Pages/AviationQuery.aspx, and these reports encompass textual narratives, findings, and recommendations from NTSB investigations, making them an invaluable resource for the study.

### B. Text Processing

Preprocessing was performed to prepare the textual data for topic modelling. The first step involved tokenization, which splits text into individual words or tokens using the NLTK library. All text was converted to lowercase for consistency, and common stopwords such as "the," "and," and "of" were removed using the NLTK stopword list. Lemmatization was applied to reduce words to their base forms, for example, converting "running" to "run," using WordNetLemmatizer. Special characters, punctuation, numbers, and other non-alphabetic characters were also filtered out during preprocessing. This process ensured that the cleaned corpus was suitable for both BERTopic and PLSA enabling a consistent comparison between the methods.

### C. *Topic Modeling Procedure*

After thorough text preprocessing the next crucial step involved transforming the preprocessed textual data into numerical representations suitable for topic modelling techniques. For this study, PLSA and BERTopic were implemented separately with topics extracted from the dataset as outlined in Fig. 1.

#### 1) *Probabilistic Latent Semantic Analysis (PLSA)*

PLSA was implemented using Gensim, a Python library for topic modelling. The document-term matrix was created to represent the text corpus as a sparse matrix, where rows correspond to documents and columns to words. The model fitting involved estimating the topic-word and document-topic distributions using the Expectation-Maximization (EM) algorithm, which then allowed for the extraction of topics as probabilistic distributions over words.

#### 2) *BERTopic*

BERTopic, in contrast, was implemented using the BERTopic library. This process began with the embedding generation, where text was converted into high-dimensional embeddings using pre-trained transformer models such as BERT. Clustering techniques, specifically HDBSCAN (Hierarchical Density-Based Spatial Clustering of Applications with Noise), were employed to group embeddings into clusters. Dynamic topic representation was then used to assign interpretable topics to clusters based on their most representative words.

### D. *Evaluation Metrics*

To compare the performance of BERTopic and PLSA, several metrics were employed. Topic coherence was a primary measure, assessing the semantic similarity between words within a topic. The coherence score was computed using the C_v metric from Gensim's Coherence Model, which correlates well with human judgment [19]. Interpretability was assessed through manual inspection by aviation safety experts focusing on the clarity and relevance of the topics. Scalability was evaluated based on the ability of each method to handle large datasets, measured in terms of computational time and memory usage.

### E. *Experimental Setup*

The experiments were conducted in a Python-based environment with a system configuration including an Intel i7 processor, 32GB RAM, and an NVIDIA GPU for transformer-based embeddings. The software environment comprised Python 3.10, BERTopic 0.13.0, Gensim 4.3.1, and NLTK 3.8.0. For PLSA, the number of topics was set to 6 based on preliminary experiments, while for BERTopic, the minimum cluster size was set to 15, with default transformer embeddings used. The results from these experiments were analyzed to determine which method is better suited for extracting key topics in aviation safety reports.

## IV. Results and Discussion

The results of this study highlight the differences between BERTopic and PLSA in extracting key topics from aviation safety reports. The comparative analysis focuses on topic coherence, interpretability, and scalability, with quantitative and qualitative findings evaluated [20]. These results are discussed below.

### A. *Topic Word Clouds*

Word clouds were generated to visually represent the most prominent words in the topics extracted by both BERTopic and PLSA, as shown in Figs. 2 and 3, respectively. For both models, larger words indicate higher importance within their respective topics. BERTopic exhibited more focused clusters of domain-specific terms, while PLSA highlighted a broader range of general terms, reflecting its probabilistic nature.

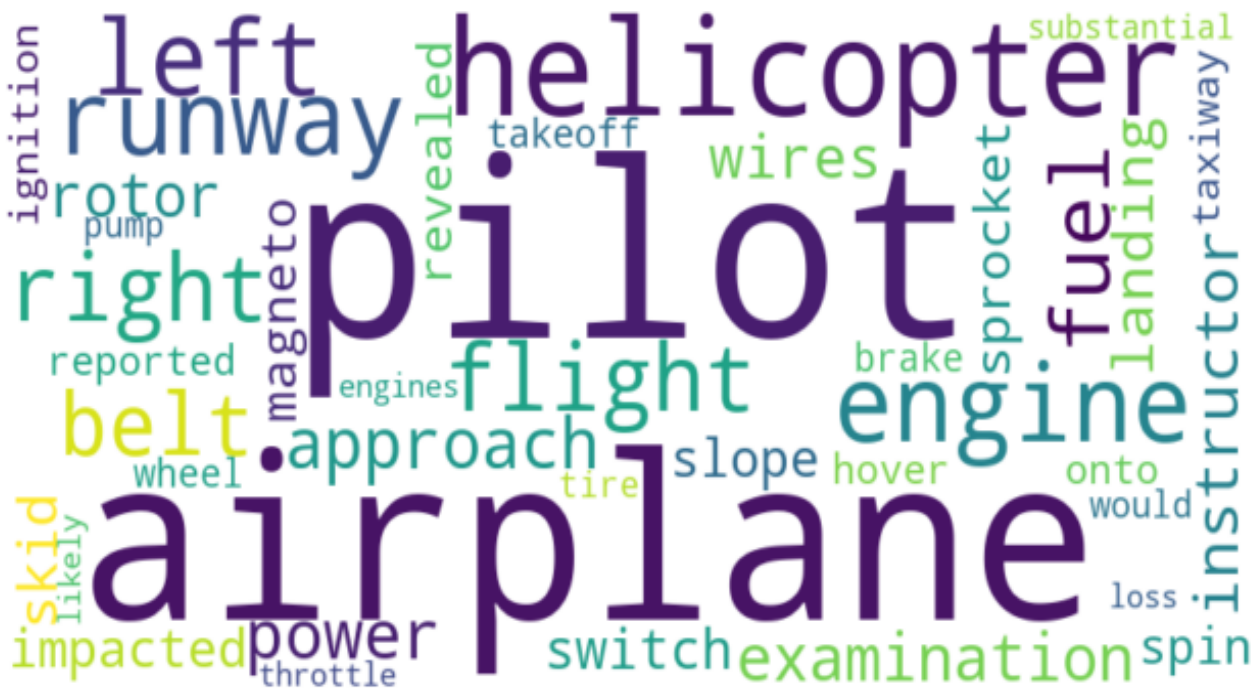

Fig. 2. Word Cloud of Topics on BERTopic Model

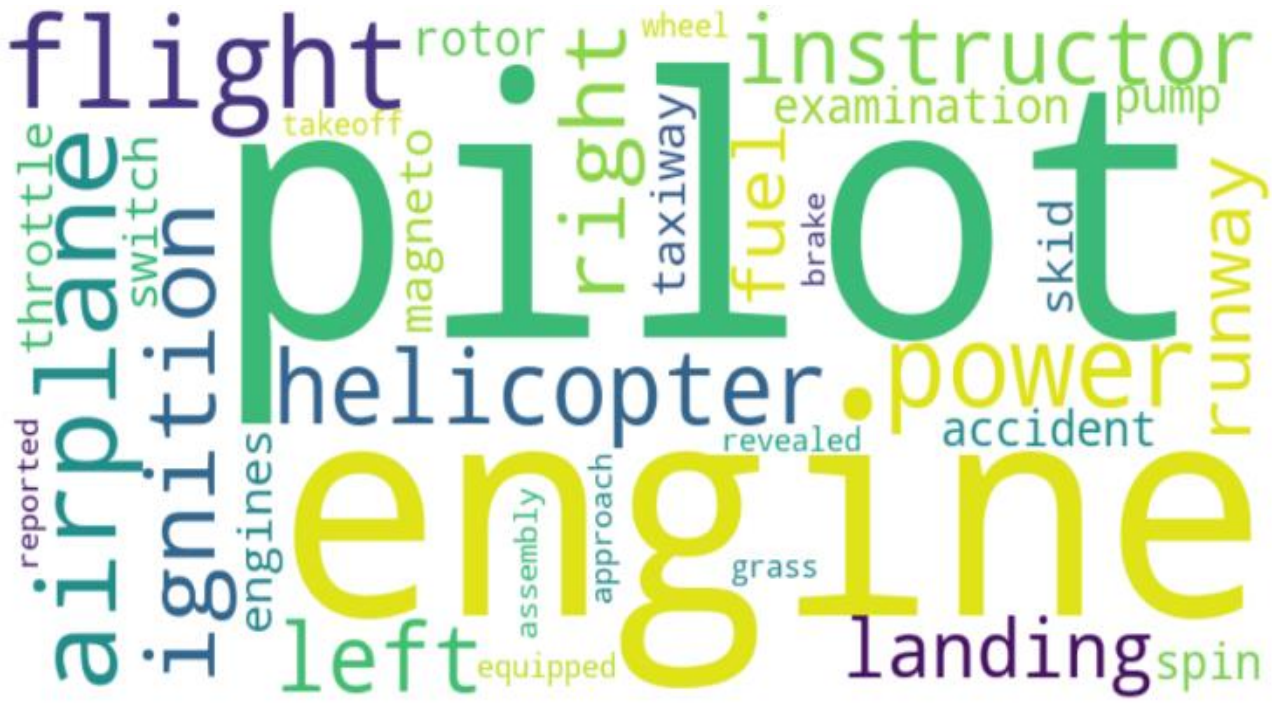

Fig. 3. Word Cloud of Topics on PLSA Model

### B. *Hierarchical Clustering for BERTopic*

Fig. 4 illustrates the hierarchical clustering of topics derived from BERTopic. Using HDBSCAN, the clustering process grouped semantically similar terms, showcasing the model's ability to identify nuanced relationships between words. This clustering aligns with the domain-specific nature of aviation safety data and enhances interpretability.

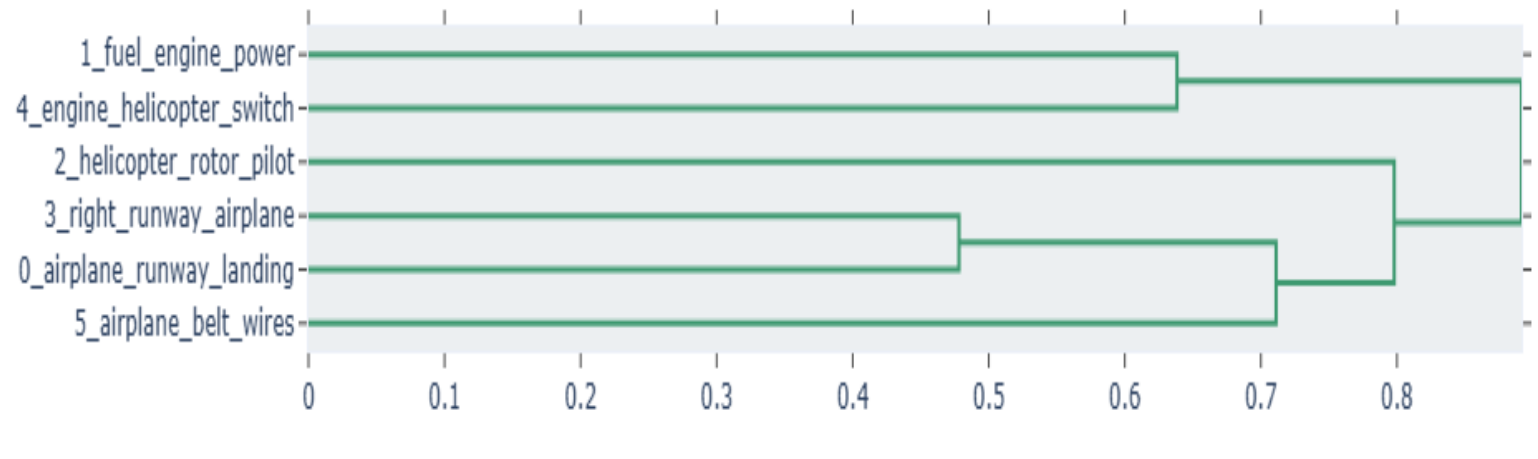

Fig. 4. Hierarchical clustering for BERTopic Model

### C. *Coherence and Interpretability*

The topic coherence scores were computed to assess the semantic similarity of words within each topic. BERTopic achieved a coherence score of 0.41, outperforming PLSA, which scored 0.37. This suggests that BERTopic's clustering approach effectively captured the contextual relationships between words, resulting in higher-quality topics.

Interpretability was evaluated by manually inspecting the topics. BERTopic demonstrated a higher degree of interpretability, with clearly defined themes such as "helicopter operations," "engine failures," and "takeoff incidents." PLSA, while identifying relevant terms, struggled to maintain semantic clarity, often blending unrelated words into the same topic, as shown in Tables I and II.

### D. *Topic Word Scores*

Figs. 5 and 6 present the word score distributions for the six topics identified by BERTopic and PLSA, respectively. These figures highlight the relative importance of individual words within each topic, with BERTopic consistently assigning higher weights to domain-specific terms.

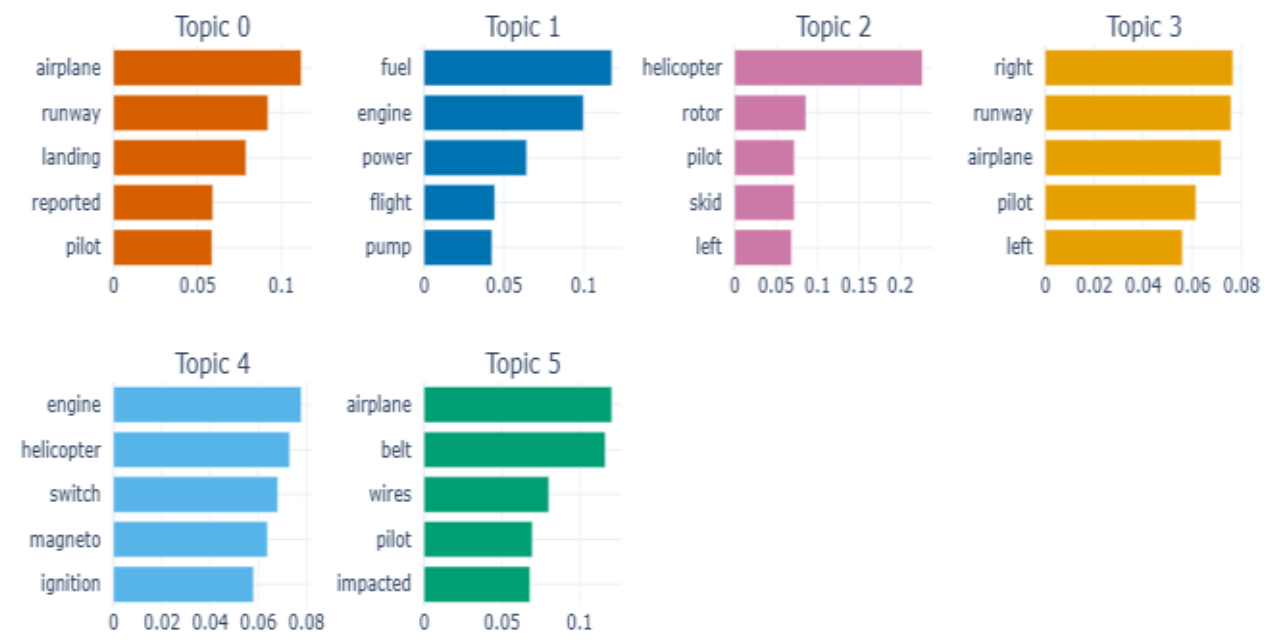

Fig. 5. Topic word score for BERTopic Model

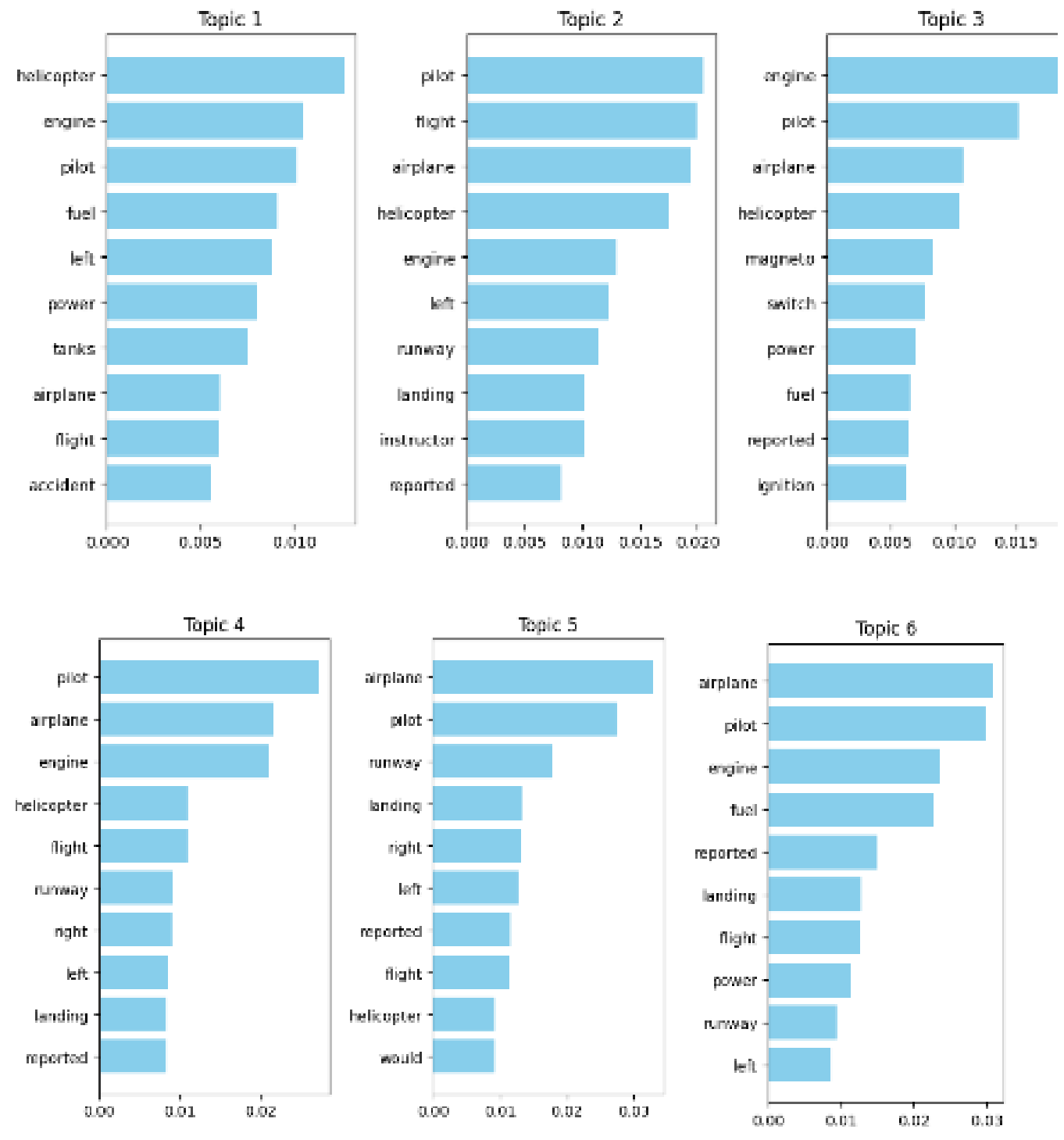

Fig. 6. Topic word score for PLSA Model

### E. *Topic Word Distributions*

Tables I and II summarize the top 10 words and the associated themes (single-word topics) for the six topics generated by BERTopic and PLSA. These themes represent the overarching subject for each topic, derived from the most representative words. While both models identified relevant terms, BERTopic demonstrated superior coherence and thematic clarity compared to PLSA.

TABLE I. BERTOPIC MODEL: TOP 10 WORDS AND THEME FOR EACH TOPIC

| Topic ID | ***Top 10 Words*** | ***Theme / Single Word*** |
|---|---|---|
| Topic 1 | airplane, runway, landing, reported, pilot, would, approach, left, flight, Substantial | Flight Operations |
| Topic 2 | fuel, engine, power, flight, pump, loss, engines, examination, pilot, likely | Fuel System and Failures |
| Topic 3 | helicopter, rotor, pilot, skid, left, spin, right, hover, onto, taxiway | Helicopter Operations |
| Topic 4 | right, runway, airplane, pilot, left, takeoff, brake, instructor, wheel, tire | Takeoff challenges and brake issues |
| Topic 5 | engine, helicopter, switch, magneto, ignition, pilot, flight, power, instructor, throttle | Engine malfunctions |
| Topic 6 | Airplane, belt, wires, pilot, impacted, sprocket, slope, approach, revealed, examination | Structural Failures |

TABLE II. PLSA MODEL: TOP 10 WORDS AND THEME FOR EACH TOPIC

| Topic ID | ***Top 10 Words*** | ***Theme / Single Word*** |
|---|---|---|
| Topic 1 | airplane, pilot, engine, fuel, runway, flight, landing, power, left, reported | Engine and Fuel systems. |
| Topic 2 | pilot, helicopter, engine, airplane, right, reported, landing, flight, switch, magneto | Flight Operations and Pilot activities |
| Topic 3 | airplane, flight, left, runway, pilot, reported, landing, would, instructor, normal | Ignition system |
| Topic 4 | engine, pilot, fuel, helicopter, power, flight, instructor, ft, left, loss | Flight Procedures |
| Topic 5 | pilot, airplane, engine, landing, helicopter, reported, runway, flight, left, normal | Runway and Airplane control |
| Topic 6 | pilot, airplane, flight, runway, reported, helicopter, engine, left, instructor, would | Engine Failure |

### F. *Comparison of Topics and Themes: BERTopic vs. PLSA*

Both BERTopic and PLSA identified overlapping themes in aviation safety, such as Flight Operations, Engine Performance, and Helicopter Mechanics. However, their approaches to grouping and interpreting these themes varied significantly. BERTopic excelled at organizing words into broader, context-driven themes, such as Flight Operations, Helicopter Operations, and Takeoff & Landing, reflecting its ability to leverage semantic relationships. This approach enabled BERTopic to capture nuanced connections between terms, providing more interpretable and holistic insights.

In contrast, PLSA focused on more specific, technical aspects, emphasizing topics like Engine Issues, Helicopter Dynamics, and Flight Procedures. Its reliance on statistical co-occurrence patterns resulted in topics that were more structured but less contextually rich. Additionally, PLSA often highlighted problem-oriented themes, such as fuel shortages and reported incidents, making it more suited for identifying discrete issues rather than broader concepts.

While both models captured similar aviation safety themes, BERTopic provided a more contextually relevant and intuitive grouping of terms, whereas PLSA offered a more statistical, detail-oriented lens. Table III summarizes the strengths and weaknesses of BERTopic and PLSA, highlighting key differences in coherence, interpretability, granularity, and scalability. It provides a clear comparison of how each model performs across various aspects of topic modelling in aviation safety reports.

*2) Word Distribution of Topics for PLSA: Fig. 8* presents the word distribution across documents for PLSA. Notably, Topic 4 emerges as the dominant topic, as it spans most of the document corpus. This indicates that PLSA tends to assign many documents to a single topic, suggesting a stronger focus on high-frequency patterns in the dataset. While this approach captures dominant themes effectively, it limits the model's ability to provide a more diverse and granular representation of topics. These visualizations highlight a key difference: BERTopic excels in generating distinct, contextually grounded topics, while PLSA emphasizes statistical dominance, often clustering many documents under fewer overarching topics.

TABLE III. COMPARISON OF TOPICS AND THEMES: BERTOPIC VS. PLSA

| Aspect | *BERTopic* | *PLSA* |
|---|---|---|
| **Topic Coherence** | High coherence due to semantic embeddings, achieving a score of 0.41. | Lower coherence due to bag-of-words representation, with a score of 0.37. |
| **Interpretability** | Superior, with clear and intuitive topics aligned with domain knowledge. | Moderate, with some overlap and less distinct topic boundaries. |
| **Granularity** | Captures fine-grained and actionable insights (e.g., specific safety issues). | Produces broader topics lacking actionable specificity. |
| **Scalability** | Efficient for large datasets; processes over 36,000 records in 1.2 hours. | Computationally intensive; takes 2.6 hours for the same dataset. |
| **Strengths** | Leverages advanced embeddings; excellent for context-rich domains. | Simpler, easier to implement, and effective for small datasets. |
| **Weaknesses** | Requires higher computational resources; embedding generation is costly. | Limited by bag-of-words approach; struggles with semantic relationships. |

## G. *Visual Analysis of BERTopic and PLSA*

To enhance the understanding of the models' performance, visualizations of topic distributions and relationships were analyzed.

*1) Intertopic Distance Map for BERTopic: Fig. 7* demonstrates the relationships between topics extracted by BERTopic. This visualization shows distinct separations between topics, indicating BERTopic's ability to cluster semantically meaningful themes effectively. The map highlights BERTopic's capability to provide contextually rich and coherent topics, which is especially beneficial in analyzing the complex aviation dataset.

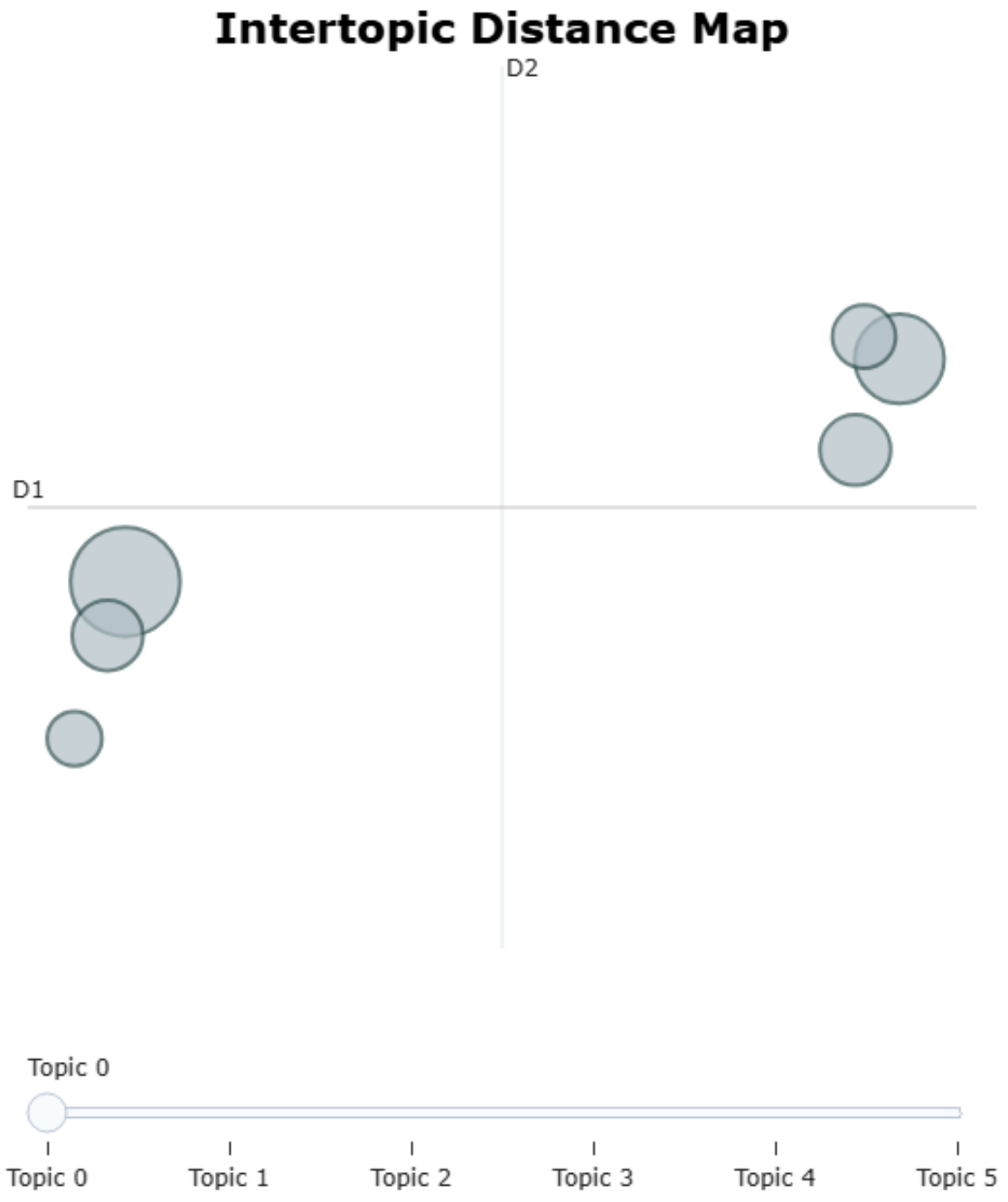

Fig. 7. BERTopic Distance Map

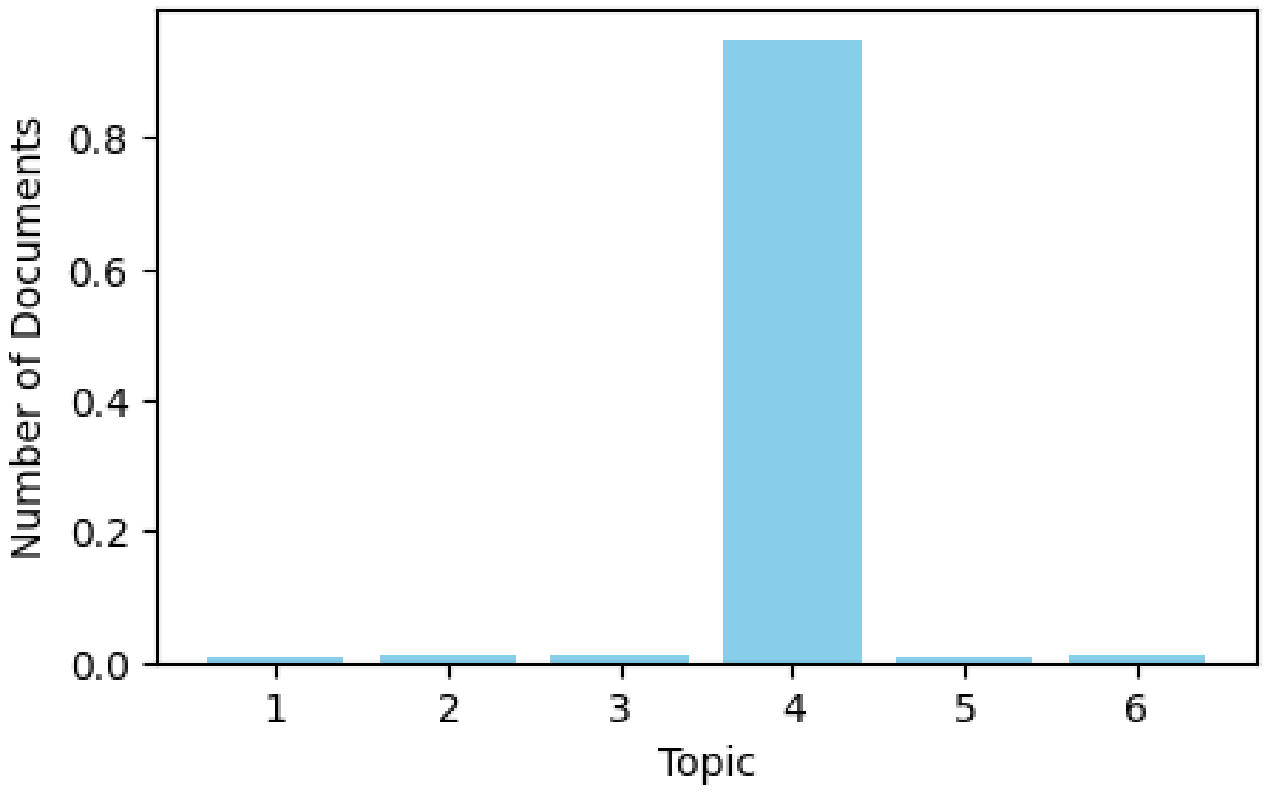

Fig. 8. PLSA document word distribution

# V. CONCLUSION

This study compared BERTopic and PLSA in extracting topics from aviation safety reports, focusing on topic coherence, interpretability, scalability, and topic-specific insights. The results demonstrated that BERTopic outperformed PLSA in several aspects. BERTopic achieved higher coherence scores (0.41 vs. 0.37) due to its ability to leverage contextual embeddings, which enhanced the semantic consistency of the topics. It also produced more interpretable and granular topics that aligned closely with domain-specific themes, such as engine failures, runway incursions, and other related incidents. In contrast, PLSA relied on a bag-of-words representation, which limited its ability to capture nuanced relationships between words. While it provided statistically sound results, its topics were less coherent and often overlapped, reducing their interpretability. Additionally, PLSA struggled with scalability, requiring significantly more computational resources and time to process large datasets than BERTopic. The findings suggest that BERTopic is better suited for analysing large-scale and complex datasets, particularly in domains like aviation safety, where contextual understanding is crucial. However, the choice of topic modelling technique should always consider the specific needs of the dataset and application. While PLSA remains a valuable method in certain scenarios, modern approaches like BERTopic, which leverage advancements in NLP, provide more actionable insights for critical domains such as aviation safety. Future research could explore hybrid approaches that combine the strengths of traditional and embedding-based models, as well as applications of these techniques in other safety-critical industries.